\documentclass[aps,prl,twocolumn,superscriptaddress,showpacs]{revtex4}
\usepackage{graphicx}
\usepackage{mathrsfs}
\usepackage{bm}
\usepackage{amsmath}
\usepackage{dcolumn}
\usepackage{epstopdf}
\usepackage{dsfont}
\usepackage{amssymb}
\usepackage{tabularx}
\usepackage{array}
\usepackage{float}
\usepackage{colordvi}

\begin{document}
\title{Current partition at zero-line intersection of quantum anomalous Hall topologies}
\author{Yafei Ren}
\affiliation{College of Physics and Energy, Shenzhen University, Shenzhen 518060, China}
\affiliation{ICQD, Hefei National Laboratory for Physical Sciences at Microscale, and Synergetic Innovation Center of Quantum Information and Quantum Physics, University of Science and Technology of China, Hefei, Anhui 230026, China}
\affiliation{CAS Key Laboratory of Strongly-Coupled Quantum Matter Physics and Department of Physics, University of Science and Technology of China, Hefei, Anhui 230026, China}
\author{Junjie Zeng}
\affiliation{ICQD, Hefei National Laboratory for Physical Sciences at Microscale, and Synergetic Innovation Center of Quantum Information and Quantum Physics, University of Science and Technology of China, Hefei, Anhui 230026, China}
\affiliation{CAS Key Laboratory of Strongly-Coupled Quantum Matter Physics and Department of Physics, University of Science and Technology of China, Hefei, Anhui 230026, China}
\author{Ke Wang}
\affiliation{ICQD, Hefei National Laboratory for Physical Sciences at Microscale, and Synergetic Innovation Center of Quantum Information and Quantum Physics, University of Science and Technology of China, Hefei, Anhui 230026, China}
\affiliation{CAS Key Laboratory of Strongly-Coupled Quantum Matter Physics and Department of Physics, University of Science and Technology of China, Hefei, Anhui 230026, China}
\author{Fuming Xu}
\thanks{xufuming@szu.edu.cn}
\affiliation{College of Physics and Energy, Shenzhen University, Shenzhen 518060, China}
\author{Zhenhua Qiao}
\thanks{qiao@ustc.edu.cn}
\affiliation{ICQD, Hefei National Laboratory for Physical Sciences at Microscale, and Synergetic Innovation Center of Quantum Information and Quantum Physics, University of Science and Technology of China, Hefei, Anhui 230026, China}
\affiliation{CAS Key Laboratory of Strongly-Coupled Quantum Matter Physics and Department of Physics, University of Science and Technology of China, Hefei, Anhui 230026, China}

\begin{abstract}
At the interface between two-dimensional materials with different topologies, topologically protected one-dimensional states (also named as zero-line modes) arise. Here, we focus on the quantum anomalous Hall effect based zero-line modes formed at the interface between regimes with different Chern numbers. We find that, these zero-line modes are chiral and unilaterally conductive due to the breaking of time-reversal invariance. For a beam splitter consisting of two intersecting zero lines, the chirality ensures that current can only be injected from two of the four terminals. Our numerical results further show that, in the absence of contact resistance, the (anti-)clockwise partitions of currents from these two terminals are the same owing to the current conservation, which effectively simplifies the partition laws. We find that the partition is robust against relative shift of Fermi energy, but can be effectively adjusted by tuning the relative magnetization strengths at different regimes or relative angles between zero lines.
\end{abstract}
\date{\today}

\maketitle

\textit{Introduction---.} The presence of edge states that are topologically protected from backscattering is one of the striking hall-marks of topologically nontrivial insulators~\cite{rev_Ren_16, rev_TI_Kane_10, rev_TI_Zhang_11, rev_QAHE_MagTI_exp_XueQK_13, rev_TBandT_Bansil_16}. According to the rigorous bulk-edge correspondence rule~\cite{Bulk_edge_correspondence_Hatsugai_93,Bulk_edge_correspondence_Porta_13}, edge states appear at the boundary of two-dimensional topological systems, like quantum Hall effect~\cite{QHE_Klitzing_04}, quantum anomalous Hall effect~\cite{QAHE_G_Qiao_10, QAHE_MagTI_FangZh_10}, and quantum spin-Hall effect (or two-dimensional topological insulators)~\cite{QSHE_G_Haldane_06, QSHE_QW_HgTe_ZhangSC_06}. These edge states are localized at the boundaries that are interfaces between topological materials and topologically trivial vacuum. Thus, these boundaries can be generalized to interfaces between two topologically distinct materials, like the interface between quantum anomalous Hall effect and quantum valley Hall effect~\cite{ZLM_G_Yang_15}, quantum spin-Hall effect and quantum valley Hall effect~\cite{ZLM_TI-QV_Yao_16}, or the graphene nanoroad between two structurally different boron-nitride sheets~\cite{ZLM_G_Nanoroad_Qiao_12}. One widely explored system is the zero-line modes (ZLMs) occurred at the interface, across which the valley Chern numbers varies~\cite{ZLM_BLG_Morpurgo_08, rev_Ren_16}. These ZLMs are protected from long-range scattering potential by large momentum separation and exhibit zero bend resistance in the absence of atomic defects~\cite{ZLM_G_Nanoroad_Qiao_12, ZLG_BLG_MacDonald_11}. Such modes are experimentally feasible~\cite{ZLM_BLG_ABBA_exp_Wang_15, ZLM_BLG_exp_Li_15} in Bernal stacked multilayer graphenes with out-of-plane electric field~\cite{QVHE_G_NiuQ_07, QSHE_BLG_Qiao_11, QSHE_TLG_Qiao_12} and have attracted much attention from both theoreticians~\cite{rev_Ren_16, ZLM_BLG_BEC_Martin_10, ZLM_G_BEC_Morpurgo_12, ZLM_MLG_NiuQ_11, ZLM_G_ZhouF_08, ZLM_G_Niu_09, ZLM_Si_Chan_14, ZLM_BLG_ABBA_Kim_13, ZLM_BLG_ABBA_Mele_13, ZLM_Si_Ihm_14, ZLM_G-BN_ring_Miller_12, ZLM_G_Xie_12, ZLG_BLG_MacDonald_11, ZLM_Analytical_Schulz_17} and experimentalists~\cite{ZLM_BLG_ABBA_exp_McEuen13, ZLM_BLG_ABBA_exp_Wang_15, ZLM_BLG_exp_Li_15}.

Comparing to the sample boundary, the interface exhibits higher tunability, e.g., one can construct few interfaces as well as intersections to form networks~\cite{ZLM_G_NiuQ_14, MxN_partition_Qiao_16}. At the intersection between zero lines of quantum valley Hall effect, counterintuitive current partition was reported~\cite{ZLM_G_NiuQ_14}, and later the tunable current partition at multiple intersections was also demonstrated~\cite{MxN_partition_Qiao_16}. In these systems, time-reversal invariance guarantees the helical transport of these ZLMs. In contrast, ZLMs based on quantum anomalous Hall effect are chiral owing to the time-reversal breaking, and are topologically protected from backscattering even in the presence of short-range atomic scatters, suggesting more interesting transport properties. 

\begin{figure*}
  \includegraphics[width=17 cm,angle=0]{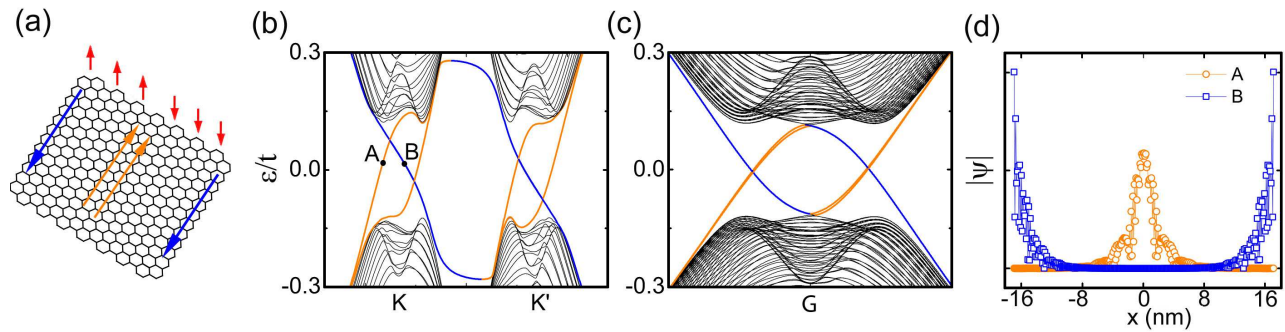}
  \caption{(a) Schematic of ZLMs from different quantum anomalous Hall topologies. The left (right) side of the sample experiences an upward (downward) exchange field as shown by red arrows. Orange arrows indicate the ZLMs while the blue ones are edge states. (b) and (c) Band structures of zigzag and armchair terminated graphene nanoribbons with Rashba SOC $t_{\rm R}/t=0.10$. Orange and blue bands are respectively the ZLMs and edge modes. (d) Normalized probability amplitudes of wave functions for states ``A" and ``B" labeled in panel (b).}\label{Band-QAHE}
\end{figure*}

In this article, we study the ZLMs formed at the interfaces between quantum anomalous Hall phases with opposite Chern numbers in graphene systems. We find that topologically protected ZLMs occur at both zigzag and armchair interfaces while the chiral edge states also appear at sample boundaries. These ZLMs are chiral and exhibit unilateral conductivity due to the breaking of time-reversal invariance, which are completely different from the ZLMs based on quantum valley Hall effect. For a four-terminal setup composed of two intersected zero lines, the chirality of these modes ensures that the current can only be injected from two of the four terminals. Further numerical calculations show that the (anti-)clockwise partitions of currents from these two different terminals are the same guaranteed by the current conservation in the absence of contact resistance. We then study the manipulation of current partition by external means, and find that, different from the ZLMs based on quantum valley Hall effect, the partition is robust against the relative shift of Fermi energy but can be effectively tuned by changing the relative magnetization strengths in different regions or the relative angles between these four terminals.

\textit{Model Hamiltonian and Band Structures---.} We employ the $\pi$-orbital tight-binding Hamiltonian of monolayer graphene with fixed Rashba spin-orbit coupling and spatially varying out-of-plane magnetization, which can be expressed as~\cite{QAHE_G_Qiao_10}:
\begin{eqnarray}
\label{EQSingleH}
H  &= & -t\sum_{\langle ij \rangle}c^\dagger_{i}c_{j} + {\mathrm{i}} t_{\mathrm{R}} \sum_{\langle{ij}\rangle} c^\dagger_{i} (\bm{s}{\times}\hat{\bm{d}}_{ij}) {\cdot} \hat{\bm{z}} \,  c_{j} \nonumber \\
& +& \sum_{i} M_i f_i c^\dagger_{i} s_z c_{i}, 
\end{eqnarray}
where $c^\dagger_{i}=(c^\dagger_{i\uparrow},c^\dagger_{i\downarrow})^{\rm{T}}$ is the creation operator for an electron at the $i$-th site with  $\uparrow$ and $\downarrow$ representing the spin-up and -down states. The first term stands for the nearest-neighbor hopping with an amplitude of $t$, and the second term is the Rashba spin-orbit coupling (SOC) involving only the nearest neighbor hopping, where $\hat{\bm{d}}_{ij}$ is a unit vector pointing from site $j$ to $i$ and $\bm{s}$ are spin-Pauli matrices. The third term represents the spatially varying out-of-plane exchange field where $M_i$ measures its strength and $f_i=\pm1$ indicates the upward (downward) orientation of the exchange field at site $i$.

In the presence of a uniform exchange field, Rashba SOC can induce a topologically nontrivial bulk band gap in monolayer graphene, hosting a quantum anomalous Hall state of Chern number $\pm2$ depending on the magnetization orientation~\cite{QAHE_G_Qiao_10}. When the exchange field varies spatially, as shown by red arrows in Fig.~\ref{Band-QAHE}(a), quantum anomalous Hall topologies of Chern numbers $\mathcal{C}=\pm2$ change between left and right sides, separately. Along the interface, unilaterally propagated ZLMs present as schematically plotted by orange arrows. The corresponding band structures of the zigzag and armchair graphene ribbons experiencing varying exchange fields are displayed in Figs.~\ref{Band-QAHE}(b) and \ref{Band-QAHE}(c), respectively. The four non-degenerate orange bands are gapless ZLMs localized at the interface while the blue bands are doubly degenerate edge modes and there are two modes at each boundary as shown in Fig.~\ref{Band-QAHE}(d). These states shall be robust against weak disorders since the counter-propagating channels are spatially separated. 

These ZLMs based on quantum anomalous Hall effect exhibit some different characters from that based on quantum valley Hall effect~\cite{ZLG_BLG_MacDonald_11}. On one hand, for armchair nanoribbons in the latter case, edge states disappear and ZLMs possesses a small topologically trivial band gap due to the strong coupling between valleys $K$ and $K'$. Here, however, both edge states and ZLMs appear for armchair nanoribbon and show rigorous gapless character due to the topological protection, which makes it possible to investigate the transport properties at the charge neutrality point. On the other hand, different from the quantum valley Hall systems where the ZLMs at $K$ and $K'$ valleys show opposite group velocities guaranteed by the time-reversal invariance, such invariance is broken in quantum anomalous Hall systems and the corresponding ZLMs propagate in the same direction for both valleys while the edge states at the boundaries propagate along opposite directions, resulting the unilateral propagation of the ZLMs.

\begin{figure}
  \includegraphics[width=6 cm, angle=0]{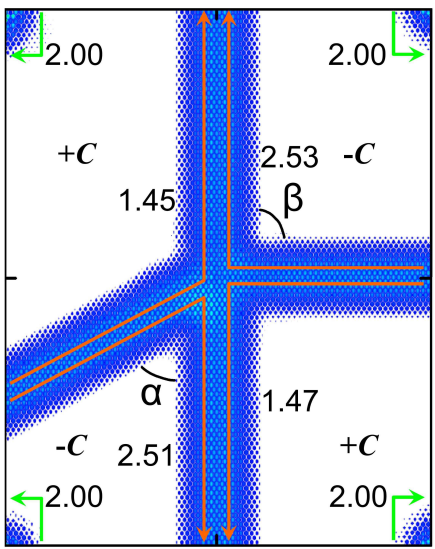}
  \caption{Fermi-level scattering state density distributions in four-probe zero-line intersection for zero-line configuration of $\alpha=60^\circ$ and $\beta=90^\circ$. The L, R, U, and D reservoirs each has a single zero line at the ribbon center which matches a system zero line at its boundary. Each point in these plots corresponds to an individual carbon atom site. The orange arrows indicate the currents flow along zero lines while the green ones at the boundaries suggest the currents along edges. The number near each orange arrow represents the currents partitioned into U or D terminals from L or R. $G_{\rm UL, DL}$ is not exactly the same as $G_{\rm DR, UR}$ due to the numerical error. $\pm\mathcal{C}=\pm 2$ stands for the Chern number at each regime.}\label{DOS-QAHE}
\end{figure}

\textit{Current partition at topological intersections---.}
We now focus on the current partition at a four-terminal device composed of two intersecting zero lines as illustrated in Fig.~\ref{DOS-QAHE}, where ZLMs appear at the interface between regimes of $\pm \mathcal{C}$ with four terminals connecting to four leads labeled as up (U), down (D), left (L), and right (R). The current partition properties are obtained by studying the conductances between these four leads. In our calculation, the zero lines connecting U and D are set to be straight, while those connecting L and R are tunable by angles of $\alpha$ (between L and D) and $\beta$ (between R and U) as shown in Fig.~\ref{DOS-QAHE}.

Our electronic transport calculations are based on the Landauer-B\"uttiker formalism~\cite{LB_formula} and recursively constructed Green's functions~\cite{GF1, GF2}. The conductance from terminal $q$ to $p$ is numerically evaluated by using $G_{pq}=(e^2/h){\rm{Tr}}[\Gamma_p G^r \Gamma_q G^a]$, where $e$ is the electron charge, $h$ is the Planck's constant, $G^{r,a}$ are the retarded and advanced Green's functions of the central scattering regime, and $\Gamma_p$ is a line-width function describing the coupling between terminal $p$ and the central region~\cite{GF1}. The current incoming from terminal $p$ and its partition to other terminals can be schematically illustrated by plotting its contribution to the local density of states at energy $\varepsilon$: $\rho_p (\bm{r},\varepsilon)=1/2\pi[G^r \Gamma_p G^a]_{\bm{rr}}$, where $\bm{r}$ is the real space coordinate. Without loss of generality, we set $\beta=90^\circ$ in the following calculations.

\textit{Unilateral transport and partition laws of ZLMs---.} Due to the chirality of these ZLMs, one may naturally expect such a four-terminal device will exhibit unilateral conductivity similar to a $p$-$n$ junction, i.e., given the configuration of magnetization orientations, the current can only be injected into the zero lines from two terminals but be output via another two terminals as illustrated in Fig.~\ref{DOS-QAHE} where the current density is denoted in color and the current can only flow from L/R to U/D terminals along the zero lines as suggested by the orange arrows. The number near each orange arrow labels the conductance of that channel. One can find that, more portion of the current from terminal L is partitioned into terminal D, when the angle between L and D is smaller than that between L and U, similar to the counterintuitive partition law in quantum valley Hall effect based zero-line intersection~\cite{ZLM_G_NiuQ_14}. For the U/D terminal, however, the current can only be injected through the edge states with quantized conductances as suggested by the green arrows. We summarize the transport rules from the numerical calculations in the following:
\begin{eqnarray}
&&G_{\rm RL}=G_{\rm LR}=G_{\rm UD}=G_{\rm DU}=0.00e^2/h; \label{equation1}\\
&&G_{\rm UL}+G_{\rm DL}=G_{\rm UR}+G_{\rm DR}=4.00e^2/h; \label{equation3}\\
&&G_{\rm RU}=G_{\rm LU}=G_{\rm RD}=G_{\rm LD}=2.00e^2/h; \label{equation2}\\
&&G_{\rm UL}=G_{\rm DR}; \label{equation4}\\
&&G_{\rm DL}=G_{\rm UR}.\label{equation5}
\end{eqnarray}
Equation~(\ref{equation1}) reflects the forbidden forward transmission due to the chirality of ZLMs. Along zero lines, the current is injected from L or R, and is partitioned into U and D terminals, giving rise to finite but unnecessarily quantized values of $G_{\rm U/D,L}$ and $G_{\rm U/D,R}$. The current conservation leads to the fact that the sum of partitioned currents to U and D terminals equals to the incident current from L or R as shown in Eq.~\eqref{equation3}. As for the current from U or D to L and R terminals, they can only propagate along the sample boundaries and quantized conductances appear as reflected by Eq.~(\ref{equation2}). These quantized values are very robust against the variation of $\alpha/\beta$, Fermi energy, etc., because the edge modes are not mixed with other states (either ZLMs or other edge modes) for large enough sample. It is noteworthy that these quantized values $G_{pq}$ are generally different from $G_{qp}$ as they transport along different paths. This is different from the current partition rules at the zero-line intersections based on quantum valley Hall effect where the edges do not contribute to conductance and $G_{qp}=G_{pq}$ is always guaranteed by the time-reversal invariance. 

Moreover, our numerical results also suggest that the clockwise (anti-clockwise) partition of currents injected from L or R terminal are the same as indicated in Eqs.~\eqref{equation4} and \eqref{equation5}, i.e., current from L to U (D) terminal is the same as that from R to D (U). It is noteworthy that in Fig.~\ref{DOS-QAHE}, $G_{\rm UL, DL}$ are not exactly the same as $G_{\rm DR, UR}$ due to the numerical error. Such feature effectively simplifies the current partition laws in such four-terminal device, making it possible to be characterized by only one parameter. These results are similar to the current partition laws in quantum valley Hall effect based zero-line intersection~\cite{ZLM_G_NiuQ_14}, which are originated from the current conservation and the relation of $G_{pq}=G_{qp}$ guaranteed by time-reversal invariance. However, as mentioned above, the breaking of time-reversal invariance in quantum anomalous Hall system invalidate the relationship of $G_{pq}=G_{qp}$. What is the physical origin of relations shown in Eqs.~\eqref{equation4} and \eqref{equation5}?

This can be attributed to the current conservation and the absence of contact resistance. In general, we can describe the partition of current from L or R by two separate parameters, i.e., the current partition for terminal L is $G_{\rm{UL}}$ and $(4-G_{\rm{UL}})$ while that for terminal R is $G_{\rm{UR}}$ and $(4-G_{\rm{UR}})$. One can consider the case of injecting currents from L and R simultaneously. In the absence of contact resistance, each of L and R terminals can inject current of $\delta_{\rm V}4e^2/h$ giving rise to a total current of $\delta_{\rm V}8e^2/h$ with $\delta_{\rm V}$ being the voltage difference between L/R and U/D terminals. These currents will be output into both U and D terminals. Since both U and D can only support no more than $\delta_{\rm V} 4e^2/h$ current, each terminal can and have to carry $\delta_{\rm V} 4e^2/h$. As a result, the current partition from R to U must be the same as that from L to D, i.e., $G_{\rm{UR}}=4-G_{\rm{UL}}=G_{\rm{DL}}$. The case is similar for $G_{\rm{DR}}$ and $G_{\rm{UL}}$. Such simple analysis can be generalized to cases when the absolute values of Chern numbers at different regimes are unequal. It is noteworthy that the above analysis is based on the assumption that there is no contact resistance between the central scattering regime and terminals. In the presence of such resistance induced by disorder, lattice mismatch, or spin polarization in leads, the partition laws in Eqs.~\eqref{equation4} and \eqref{equation5} are no longer exactly valid.

\begin{figure}
	\includegraphics[width=8.5 cm, angle=0]{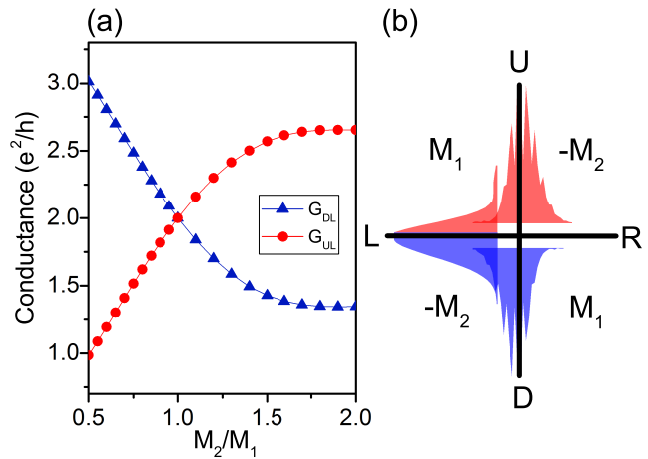}
	\caption{(a) Conductances $\rm G_{UL}$ and $\rm G_{DL}$ \textit{v.s.} relative exchange field strength $M_2/M_1$ at a fixed structure with $\alpha=90^\circ$ at the charge neutrality point. (b) Schematic of probability amplitudes of ZLMs at different zero lines with $M_2/M_1=0.5.$ Here, we set $M_1=0.2t$. More overlap between ZLMs on L and D appears.}\label{relativeMstrength}
\end{figure}

\textit{Tuning current partition via magnetization---.} We fix the system structures by setting $\alpha=\beta=90^\circ$ to investigate the role of relative exchange field strengths between neighboring quadrants. The Rashba SOC strength is fixed at $t_{\rm{R}}=0.1t$. We denote the exchange fields in the top-left, top-right, bottom-left, and bottom-right quadrants as $(+M_1,-M_2,-M_2,+M_1)$ with $M_1=0.2t$ here and $\pm$ indicating upward or downward orientation. By changing the magnitude of $M_2$, one can investigate the dependence of current partition on $M_2/M_1$ as shown in Fig.~\ref{relativeMstrength}. One can find that half-half current partition appears at $M_2=M_1$. When $M_2< M_1$, one can find that more current from L is partitioned into the D. When $M_2>M_1$, the trend of current partition becomes opposite. Thus, one can conclude that the current prefers to flow around the region with smaller magnetization.

Such findings can be understood from the dependence of wave functions of ZLMs on magnetization $M$. In graphene system, the presence of out-of-plane magnetization splits the spin-degeneracy of energy bands and leads to band crossing around Dirac points. The further introduction of Rashba SOC lift the degeneracies of band crossings to form band gaps harboring quantum anomalous Hall effect~\cite{QAHE_G_Qiao_10}. For relatively small Rashba SOC, the band gap is thus proportional to the strength of Rashba SOC, which shows nearly linear dependence on momentum. Therefore, as the magnetization strength increases, a larger band gap is expected. Generally, a larger band gap indicates a stronger topological confinement leading to smaller wave function extension as shown in Fig.~\ref{relativeMstrength}(b) where the magnetization strength is smaller in the bottom-left and top-right regimes where the ZLMs are more extended. As a result, at the zero-line intersection, the wave functions of ZLMs on L show more overlap with that on D leading to a larger current partition from L to D.

\begin{figure}
	\includegraphics[width=8.4 cm,angle=0]{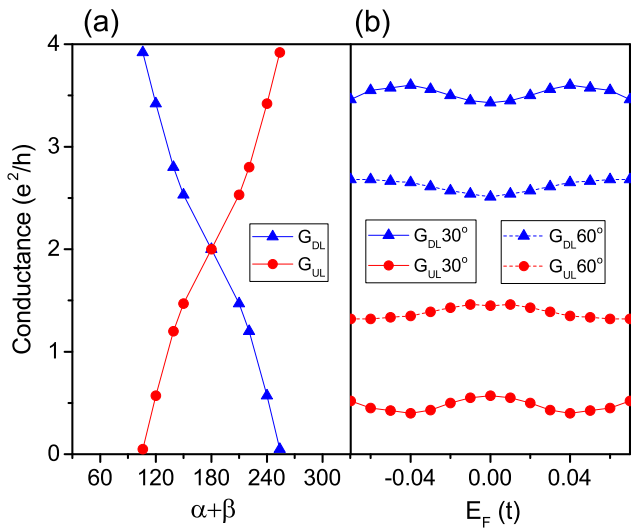}
	\caption{(a) Conductances $G_{\rm UL}$ and $G_{\rm DL}$ \textit{v.s.} $\alpha+\beta$ at the charge neutrality point $E_F=0$ with $\beta=90^\circ$. (b) $G_{\rm UL}$ and $G_{\rm DL}$ \textit{v.s.} Fermi energy $E_F$ for a fixed structure with $\beta=90^\circ$ and $\alpha=30^\circ$ or $60^\circ$. }\label{G-Angle}
\end{figure}

\textit{Tuning current partition via angle of incidence---.} We then study the current partition for different $\alpha$ at fixed $\beta=90^\circ$. By setting the Fermi energy to be zero, we calculate the conductances between different terminals and plot $G_{\rm{UL}}$ and $G_{\rm{DL}}$ as function of $\alpha+\beta$ in Fig.~\ref{G-Angle}(a). One can find that the current partition exhibits strong dependence on $\alpha$. To be specific, when $\alpha<16.1^\circ$, the left zero line is much closer to the down one and thus the ZLMs along these two zero lines show larger spatial overlap, making almost all the currents be partitioned from L to D. As $\alpha$ increases, the currents partitioned to U gradually increase, and the equal current partition occurs at $\alpha = 90^\circ$ that corresponds to a highly symmetric structure. It is noteworthy that when $\beta\neq 90^\circ$, the condition of equal partition can be generalized to be $\alpha+\beta=180^\circ$. In this case, the system is invariant under mirror reflection about the vertical zero line, which guarantees $G_{\rm DL}=G_{\rm DR}$. By combining with Eq.~\eqref{equation4}, one can find that $G_{\rm DL}=G_{\rm UL}$. As $\alpha$ further increases, more currents are partitioned into U. These behaviors are qualitatively similar to that of zero-line intersection based on quantum valley Hall effect reported in Ref.~[\onlinecite{ZLM_G_NiuQ_14}], while quantitatively different as the dependence of current partition on $\alpha+\beta$ deviates from sinusoidal function.

\textit{Tuning current partition via Fermi level---.} Here, we explore the influence of Fermi energy at fixed structures with $\alpha=30^\circ$ or $60^\circ$ as displayed in Fig.~\ref{G-Angle}(b) in solid and dashed lines, respectively. One can find that, different from the strong dependence on the angle of incidence $\alpha$, the current partition is quite robust against the Fermi energy. The relative variation of $G_{\rm UL}$ ($G_{\rm DL}$) is around $10\%$ when the Fermi level approaches the conduction band minimal or valence band maximal. Such dependence of current partition on Fermi energy is completely distinct from that in the zero-line intersection based on quantum valley Hall effect. On one hand, the system is gapped at the charge neutrality point for the latter case as the armchair boundary involves. On the other hand, the partition of current rapidly becomes equal as the Fermi energy approaches the band extrema~\cite{ZLM_G_NiuQ_14}.

\textit{Summary---.} We have numerically studied the electronic structures and wave functions of ZLMs from spatially varying quantum anomalous Hall topologies. Both ZLMs and edge states exhibit gapless character due to topological protection. The breaking of time-reversal invariance and Chern numbers of opposite signs at two sides lead to unilaterally propagating ZLMs. At the intersection of such zero lines, we have further investigated the partition laws of current flowing along zero lines. The chirality of quantum anomalous Hall effect makes the current along zero lines can only be injected from two of the four terminals (e.g., L and R) while be output from the other two (e.g., U and D). In the absence of contact resistance, even though the time-reversal invariance is absent, the current conservation guarantees the equivalence between the clockwise partitions of currents from L and R terminals in the absence of contact resistance, which greatly simplifies the current partition law to be determined by only one parameter. Different from the quantum valley Hall based topological intersection, the current partition is robust against the variation of Fermi energy. We further show that the current partition is externally tunable. By changing the relative angles between different zero lines, we find that the partition of current prefers a sharp turn. Moreover, the current partition can also be effectively tuned by changing the relative magnetization strengths at different regions and the current prefers to flow around the regions with smaller magnetization strength. Such a tunability indicates potential applications in low-energy consumption electronics. By making use of this current splitter, we can obtain tunable current propagating from L/R to U/D terminals along zero lines, and robust current propagating from U/D to L/R terminals along edges. In this way, one can fabricate tunable electronics which can produce tunable and constant current streams. Experimentally, the quantum anomalous Hall effect is proposed to appear in graphene with insulating (anti-)ferromagnetic substrate~\cite{QAHE_AFM_Qiao_14}. Thus, the ZLMs and their intersections from quantum anomalous Hall topologies are naturally expected at the domain walls of the magnetic substrate. 

\textit{Acknowledgments---.} Y.F.R. thanks for the support during his visit at Shenzhen University. This work was financially supported by the National Key R \& D Program (2016YFA0301700), NNSFC (11474265, 11504240), the China Government Youth 1000-Plan Talent Program. The Supercomputing Center of USTC is gratefully acknowledged for the high-performance computing assistance.

\end{document}